\documentclass[aps,prl, showpacs,onecolumn]{revtex4}
\usepackage{amssymb}
\usepackage{amsmath}
\usepackage{graphicx}
\usepackage{epsfig}

\begin{document}

\title{Quantum entanglement of two flux qubits induced by an auxiliary SQUID}
\author{T. \surname{Shi}}
\author{Z. \surname{Song}}
\email{songtc@nankai.edu.cn}
\affiliation{Department of Physics, Nankai University, Tianjin 300071, China}

\begin{abstract}
We revisit a theoretical scheme to create quantum entanglement of two
three-levels superconducting quantum interference devices (SQUIDs) with the
help of an auxiliary SQUID. In this scenario, two three-levels systems are
coupled to a quantized cavity field and a classical external field and thus
form dark states. The quantum entanglement can be produced by a quantum
measurement on the auxiliary SQUID. Our investigation emphasizes the quantum
effect of the auxiliary SQUID. For the experimental feasibility and
accessibility of the scheme, we calculate the time evolution of the whole
system including the auxiliary SQUID. To ensure the efficiency of generating
quantum entanglement, relations between the measurement time and dominate
parameters of the system are analyzed according to detailed calculations.

\end{abstract}

\maketitle

\section{I. Introduction}

Recently, superconducting quantum circuits (SQCs) based on superconducting
quantum interference devices (SQUIDs) \cite{Makhlin,Nakamura,Rouse,Han,Chu},
have appeared to be among the most promising candidates for quantum
information processing. Especially, schemes of quantum information transfer
(QIT) and quantum entanglement by using SQC in a microwave cavity are
carried out in many experiments \cite{Yang,Steinbach,Yu,Berkley}. The
essential idea beyond these schemes is listed as follows. Firstly, one
should make several SQUIDs \cite{Shnirman} coupling to a single mode quantum
field in a microwave cavity. Secondly, QIT in the quantum computing can be
realized by adjusting the external field to carry out a sequence of quantum
logic operations on SQUIDs.

In this aspect, there are many experiments by using SQUIDs \cite%
{Yang,Steinbach} to realize quantum entanglement of macroscopic quantum
devices. Here, a theoretical protocol has been proposed based on the
experimentally accessible parameters \cite{SI,Yang,Zhou,ZE}, in which two
SQUIDs are placed into a microwave cavity. Each flux qubit has a $\Lambda $%
-type energy level configuration. An external classical field is applied to
this system. Because these two three-level artificial atoms (or SQUIDs)
couple to a single-mode quantum field simultaneously, an effective coupling
between two three-level systems are induced by this single mode field. This
field behaves as a data bus to link two qubits coherently. When two
three-level systems coupling to two fields, a classical field and a
quantized cavity mode, there is usually a dark state, which contains a
component with maximal entanglement of two qubits. To realize the maximal
quantum entanglement\ between two qubits via the dark state, an auxiliary
SQUID is used to carry out a projective quantum measurement. As a result of
post selection measurement by wave function collapse, a quantum state with
maximal entanglement can be obtained.

Although this scheme is very subtle in the theoretical setup, there are
still some further considerations of physical mechanism needed. Especially,
the influence of the auxiliary SQUID should be taken into account since it
should be treated as a quantum subsystem. According to the contemporary
theory of quantum measurement, apparatus are usually considered as a quantum
subsystem interacting with the measured system. In fact, the quantum effect
of an auxiliary SQUID in Kansas scheme \cite{Yang} is not considered in
details. In this paper, we will stress the quantum effect of the auxiliary
SQUID with respect to the whole quantum system. Our result shows that the
back action of the auxiliary SQUID on two qubits cannot be ignored.

\begin{figure}[tbp]
\hspace{24pt}\includegraphics[bb=237 346 590 744, width=9
cm,clip]{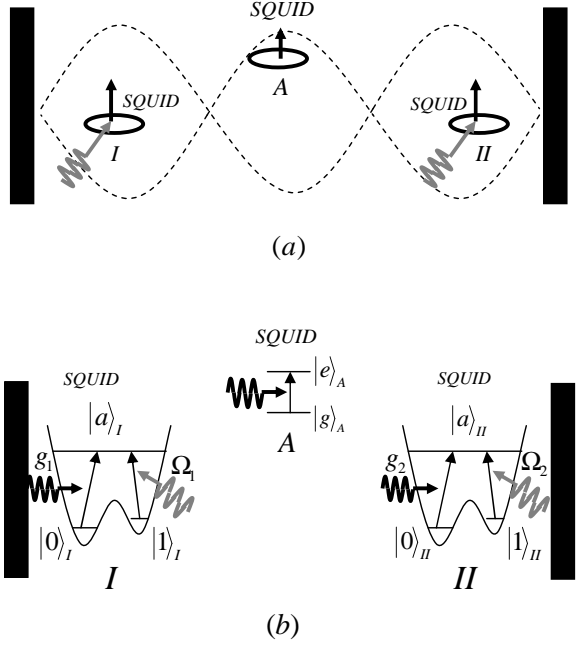}
\caption{\textit{Schematic illustration of the setup. (a) Two SQUIDs and a
auxiliary SQUID are put in a microwave cavity. Here, the black line (dashed
lines) denotes single mode cavity quantum field and grey lines denote
microwave classical fields. (b) Black lines denote single mode cavity
quantum fields. Grey lines denote microwave classical fields. Coupling
constants between the cavity field and two SQUIDs are $g_{l}$ ($l=1,2$).
Coupling constants between the external microwave field and two SQUIDs are $%
\Omega _{l}$ ($l=1,2 $).}}
\end{figure}

\section{II. The General description of quantum entanglement scheme of SQUIDs%
}

This paper is based on the theoretical proposal of Ref. \cite{Yang}. We
first briefly review the essential idea of this paper with the setup being
illustrated in Fig. 1. In each SQUID ($I$ and $II$) the transition from the
ground state $\left\vert 0\right\rangle _{l}$ to the excited state $%
\left\vert a\right\rangle _{l}$ resonates with the same single-mode cavity
field. The transition from the first excited state $\left\vert
1\right\rangle _{l}$ to the excited state $\left\vert a\right\rangle _{l}$
resonates with a external classical field. The Hamiltonian $H_{0}$ \cite%
{Yang,You,Nori} is written in the interaction picture as%
\begin{equation}
H_{0}=\sum_{l=I,II}(g_{l}c\left\vert a\right\rangle _{l}\left\langle
0\right\vert +\Omega _{l}\left\vert a\right\rangle _{l}\left\langle
1\right\vert +H.c.)\text{,}  \label{Hamiltonian}
\end{equation}%
where $g_{l}$ $\left( l=I,II\right) $ are coupling constants between the
cavity field and two SQUIDs respectively, $\Omega _{l}$ $\left(
l=I,II\right) $ are coupling constants between the external field and the
two SQUIDs respectively, $c^{\dagger }$ and $c$ are creation and
annihilation operators of the cavity field mode. In Ref. \cite{Yang}, $%
\Omega _{l}$ is time dependent to investigate the dynamics of a quantum
information transfer between two SQUIDs. In this paper we only concern the
quantum entanglement of SQUIDs and the effect of the auxiliary SQUID. So $%
\Omega _{l}$\ is a constant throught this paper.

For the Hamiltonian (\ref{Hamiltonian}), the total excitation number
\begin{equation}
\mathcal{N}_{0}=\sum_{l}\left( \left\vert a\right\rangle _{l}\left\langle
a\right\vert +\left\vert 1\right\rangle _{l}\left\langle 1\right\vert
\right) +c^{\dagger }c,
\end{equation}%
is conserved, i.e., $[H_{0},\mathcal{N}_{0}]=0$. The Hamiltonian can be
diagonalized in each invariant subspace. In this paper, we focus on $%
\mathcal{N}_{0}=1$\ subspace, which is spanned by the following basis%
\begin{eqnarray}
\left\vert \psi _{1}\right\rangle &=&\left\vert 0\right\rangle
_{I}\left\vert 0\right\rangle _{II}\left\vert 1\right\rangle _{c}\text{,}
\notag \\
\left\vert \psi _{2}\right\rangle &=&\left\vert a\right\rangle
_{I}\left\vert 0\right\rangle _{II}\left\vert 0\right\rangle _{c}\text{,}
\notag \\
\left\vert \psi _{3}\right\rangle &=&\left\vert 0\right\rangle
_{I}\left\vert a\right\rangle _{II}\left\vert 0\right\rangle _{c}\text{,}
\notag \\
\left\vert \psi _{4}\right\rangle &=&\left\vert 1\right\rangle
_{I}\left\vert 0\right\rangle _{II}\left\vert 0\right\rangle _{c}\text{,}
\notag \\
\left\vert \psi _{5}\right\rangle &=&\left\vert 0\right\rangle
_{I}\left\vert 1\right\rangle _{II}\left\vert 0\right\rangle _{c}\text{,}
\end{eqnarray}%
where $\left\vert 1\right\rangle _{c}$\ and $\left\vert 0\right\rangle _{c}$%
\ denote cavity mode states with $1$ and $0$ photon number respectively.
Diagonalizing the matrix of $H_{0}$ in this subspace we obtain the eigenstate%
\begin{eqnarray}
\left\vert d\right\rangle &=&N(\Omega _{2}g_{1}\left\vert 1\right\rangle
_{I}\left\vert 0\right\rangle _{II}\left\vert 0\right\rangle _{c}+\Omega
_{1}g_{2}\left\vert 0\right\rangle _{I}\left\vert 1\right\rangle
_{II}\left\vert 0\right\rangle _{c}  \label{dark} \\
&&-\Omega _{1}\Omega _{2}\left\vert 0\right\rangle _{I}\left\vert
0\right\rangle _{II}\left\vert 1\right\rangle _{c})\text{,}  \notag
\end{eqnarray}%
with vanishing eigenvalue. Here, $N=[(\Omega _{2}g_{1})^{2}+(\Omega
_{1}g_{2})^{2}+(\Omega _{1}\Omega _{2})^{2}]^{-1/2}$ is the normalized
factor. This is the so-called dark state because it does not couple to
excited states $\left\vert a\right\rangle _{l}$. When a photon is detected,
this state is collapsed to state $\Omega _{2}g_{1}\left\vert 1\right\rangle
_{I}\left\vert 0\right\rangle _{II}\left\vert 0\right\rangle _{c}+\Omega
_{1}g_{2}\left\vert 0\right\rangle _{I}\left\vert 1\right\rangle
_{II}\left\vert 0\right\rangle _{c}$, which is the maximally entangled state
in the case $\Omega _{2}g_{1}=\Omega _{1}g_{2}$. This is the main mechanism
for the generation of entanglement. In Ref. \cite{Yang}, an auxiliary SQUID
is employed to probe the state of the cavity. In this scheme the state of
cavity field is measured by observing the state of the auxiliary SQUID.

This task can be accomplished in the following process. A two-level
auxiliary SQUID (see Fig. 2) can be prepared initially in the ground state $%
\left\vert g\right\rangle _{A}$. It is obvious that, if the cavity field is
initially in a single photon state $\left\vert 1\right\rangle _{c}$ the
auxiliary SQUID will evolve into the excited state $\left\vert
e\right\rangle _{A}$ after a half period of Rabi oscillation; if the cavity
field is in the photon state $\left\vert 0\right\rangle _{c}$ the auxiliary
SQUID always keeps in the ground state $\left\vert g\right\rangle _{A}$.
Therefore, after a half period of Rabi oscillation, one can measure the
auxiliary SQUID to detect the cavity indirectly. If the auxiliary SQUID
collapses to the ground state $\left\vert g\right\rangle _{A}$, the cavity
field collapses to the corresponding state $\left\vert 0\right\rangle _{c}$.
Then an entangled state of two SQUIDs%
\begin{equation}
\left\vert G\right\rangle =M\left[ \Omega _{2}g_{1}\left\vert 1\right\rangle
_{I}\left\vert 0\right\rangle _{II}+\Omega _{1}g_{2}\left\vert
0\right\rangle _{I}\left\vert 1\right\rangle _{II}\right] \text{,}
\label{entangled}
\end{equation}%
is created with $M=[(\Omega _{2}g_{1})^{2}+(\Omega _{1}g_{2})^{2}]^{-1/2}$
being the normalized factor.
\begin{figure}[tbp]
\includegraphics[bb=221 522 452 684, width=7 cm, clip]{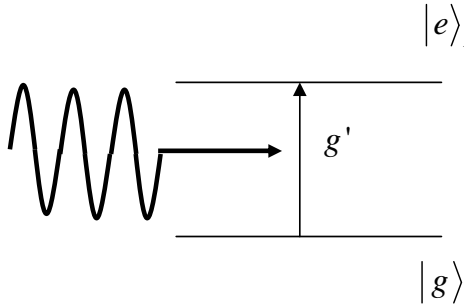}
\caption{\textit{A two-level system illustrating an auxiliary SQUID. The
coupling constant between the cavity field and auxiliary SQUID is $g^{\prime
}$.}}
\end{figure}

This the central idea of Kansas scheme, from which we can see that a
post-selection measurement plays a key role in creating the flux qubit
entanglement. The post-selection measurement is performed by an auxiliary
SQUID. In the following, we consider the auxiliary SQUID as a part of whole
quantum system.

\section{III. Dynamical evolution}

In this section the auxiliary SQUID will be taken into account as a quantum
subsystem. And the dynamical evolution of the measured system (two SQUIDs)
will be considered. Together with the auxiliary SQUID, the whole system
contains two parts, the auxiliary SQUID as the apparatus and the two SQUIDs
as the measured system. Therefore, in the interaction picture, the
Hamiltonian of the whole system reads as%
\begin{eqnarray}
H &=&H_{0}+H_{A}\text{,}  \label{whole H} \\
H_{A} &=&g^{\prime }c\left\vert e\right\rangle _{A}\left\langle g\right\vert
+H.c.\text{.}  \notag
\end{eqnarray}%
$H_{A}$ describes the interaction between the auxiliary SQUID and the cavity
field with the coupling constant $g^{\prime }$. For simplicity, in the
following we consider the case with the parameters%
\begin{eqnarray}
g_{1} &=&g_{2}=g\text{,}  \notag \\
\Omega _{1} &=&\Omega _{2}=\Omega \text{.}
\end{eqnarray}%
Obviously, we have
\begin{equation}
\lbrack H,\mathcal{N}]=0,
\end{equation}%
where\
\begin{equation}
\mathcal{N}=\sum_{l}\left( \left\vert a\right\rangle _{l}\left\langle
a\right\vert +\left\vert 1\right\rangle _{l}\left\langle 1\right\vert
\right) +\left\vert a\right\rangle _{A}\left\langle a\right\vert +c^{\dagger
}c.
\end{equation}%
When $g^{\prime }$ is switched off, the dark state $\left\vert
d\right\rangle $ from (\ref{dark}) does not evolve since it is an eigenstate
of $H_{0}$. When the auxiliary SQUID is employed to measure the state of the
cavity field, $g^{\prime }$ is switched on,\ state $\left\vert
d\right\rangle $ will evolve driven by the Hamiltonian (\ref{whole H}). If
the auxiliary SQUID is in the ground state $\left\vert g\right\rangle _{A}$
initially, while two-SQUID system is in the state $\left\vert d\right\rangle
$, the initial state will evolve in the invariant subspace $V_{s}$\ with $%
\mathcal{N}=1$,\ which is spanned by the following basis%
\begin{eqnarray}
\left\vert \varphi _{1}\right\rangle &=&\left\vert 0\right\rangle
_{I}\left\vert 0\right\rangle _{II}\left\vert 1\right\rangle _{c}\left\vert
g\right\rangle _{A}\text{,}  \notag \\
\left\vert \varphi _{2}\right\rangle &=&\left\vert a\right\rangle
_{I}\left\vert 0\right\rangle _{II}\left\vert 0\right\rangle _{c}\left\vert
g\right\rangle _{A}\text{,}  \notag \\
\left\vert \varphi _{3}\right\rangle &=&\left\vert 0\right\rangle
_{I}\left\vert a\right\rangle _{II}\left\vert 0\right\rangle _{c}\left\vert
g\right\rangle _{A}\text{,} \\
\left\vert \varphi _{4}\right\rangle &=&\left\vert 0\right\rangle
_{I}\left\vert 0\right\rangle _{II}\left\vert 0\right\rangle _{c}\left\vert
e\right\rangle _{A}\text{,}  \notag \\
\left\vert \varphi _{5}\right\rangle &=&\left\vert 1\right\rangle
_{I}\left\vert 0\right\rangle _{II}\left\vert 0\right\rangle _{c}\left\vert
g\right\rangle _{A}\text{,}  \notag \\
\left\vert \varphi _{6}\right\rangle &=&\left\vert 0\right\rangle
_{I}\left\vert 1\right\rangle _{II}\left\vert 0\right\rangle _{c}\left\vert
g\right\rangle _{A}\text{.}  \notag
\end{eqnarray}%
The corresponding matrix of $H$ in the subspace $V_{s}$ is%
\begin{equation}
H=\left(
\begin{array}{cccccc}
0 & g & g & g^{\prime } & 0 & 0 \\
g & 0 & 0 & 0 & \Omega & 0 \\
g & 0 & 0 & 0 & 0 & \Omega \\
g^{\prime } & 0 & 0 & 0 & 0 & 0 \\
0 & \Omega & 0 & 0 & 0 & 0 \\
0 & 0 & \Omega & 0 & 0 & 0%
\end{array}%
\right) \text{.}
\end{equation}%
Considering the symmetry the whole system due to the identity of two SQUIDs (%
$I$ and $II$), the $6\times 6$ matrix can be block-diagonalized under the
new basis
\begin{eqnarray}
\left\vert \chi _{1}^{\left( -\right) }\right\rangle &=&\frac{1}{\sqrt{2}}%
\left( \left\vert \varphi _{2}\right\rangle -\left\vert \varphi
_{3}\right\rangle \right) \text{,}  \notag \\
\left\vert \chi _{2}^{\left( -\right) }\right\rangle &=&\frac{1}{\sqrt{2}}%
\left( \left\vert \varphi _{5}\right\rangle -\left\vert \varphi
_{6}\right\rangle \right) \text{,}  \notag \\
\left\vert \chi _{1}^{\left( +\right) }\right\rangle &=&\left\vert \varphi
_{1}\right\rangle \text{,} \\
\left\vert \chi _{2}^{\left( +\right) }\right\rangle &=&\frac{1}{\sqrt{2}}%
\left( \left\vert \varphi _{2}\right\rangle +\left\vert \varphi
_{3}\right\rangle \right) \text{,}  \notag \\
\left\vert \chi _{3}^{\left( +\right) }\right\rangle &=&\left\vert \varphi
_{4}\right\rangle \text{,}  \notag \\
\left\vert \chi _{4}^{\left( +\right) }\right\rangle &=&\frac{1}{\sqrt{2}}%
\left( \left\vert \varphi _{5}\right\rangle +\left\vert \varphi
_{6}\right\rangle \right) \text{.}  \notag
\end{eqnarray}%
The matrix of the Hamiltonian $H$ can be written as%
\begin{equation}
H=H_{2}\oplus H_{4}\text{,}
\end{equation}%
where
\begin{equation}
H_{2}=\left(
\begin{array}{cc}
0 & 1 \\
1 & 0%
\end{array}%
\right) \text{,}
\end{equation}%
and%
\begin{equation}
H_{4}=\left(
\begin{array}{cccc}
0 & \sqrt{2}g & g^{\prime } & 0 \\
\sqrt{2}g & 0 & 0 & \Omega \\
g^{\prime } & 0 & 0 & 0 \\
0 & \Omega & 0 & 0%
\end{array}%
\right) \text{.}
\end{equation}%
Here, $g$ and $g^{\prime }$ are rescaled by $\Omega $: $g/\Omega \rightarrow
g$, $g^{\prime }/\Omega \rightarrow g^{\prime }$.

Diagonalizing two matrices, eigenvalues are obtained as%
\begin{eqnarray}
E_{1}^{\left( -\right) } &=&-1\text{,}  \notag \\
E_{2}^{\left( -\right) } &=&1\text{,}  \notag \\
E_{1}^{\left( +\right) } &=&\frac{1}{\sqrt{2}}[\sqrt{\eta ^{2}-4g^{\prime 2}}%
+\eta ]^{\frac{1}{2}}\text{,} \\
E_{3}^{\left( +\right) } &=&\frac{1}{\sqrt{2}}[-\sqrt{\eta ^{2}-4g^{\prime 2}%
}+\eta ]^{\frac{1}{2}}\text{,}  \notag \\
E_{2}^{\left( +\right) } &=&-E_{1}\text{,}  \notag \\
E_{4}^{\left( +\right) } &=&-E_{3}\text{,}  \notag
\end{eqnarray}%
where%
\begin{equation}
\eta =1+2g^{2}+g^{\prime 2}.
\end{equation}

In the antisymmetry basis: $\{\left\vert \chi _{n}^{\left( -\right)
}\right\rangle \left\vert \text{ }n=1,2\right. \}$ and symmetry basis: $%
\{\left\vert \chi _{n}^{\left( +\right) }\right\rangle \left\vert \text{ }%
n=1,2,3,4\right. \}$, corresponding eigenstates are expressed as%
\begin{eqnarray}
\left\vert \Psi _{1}^{\left( -\right) }\right\rangle &=&\frac{1}{\sqrt{2}}%
\left( \left\vert \chi _{1}^{\left( -\right) }\right\rangle -\left\vert \chi
_{2}^{\left( -\right) }\right\rangle \right) \text{,}  \notag \\
\left\vert \Psi _{2}^{\left( -\right) }\right\rangle &=&\frac{1}{\sqrt{2}}%
\left( \left\vert \chi _{1}^{\left( -\right) }\right\rangle +\left\vert \chi
_{2}^{\left( -\right) }\right\rangle \right) \text{,}  \notag \\
\left\vert \Psi _{1}^{\left( +\right) }\right\rangle &=&N_{1}(\frac{%
E_{3}^{2}-1}{g}\left\vert \chi _{1}^{\left( +\right) }\right\rangle +\sqrt{2}%
E_{3}\left\vert \chi _{2}^{\left( +\right) }\right\rangle -\frac{E_{3}}{%
g^{\prime }}\frac{E_{5}^{2}-g^{\prime 2}}{g}\left\vert \chi _{3}^{\left(
+\right) }\right\rangle +\sqrt{2}\left\vert \chi _{4}^{\left( +\right)
}\right\rangle )\text{,}  \notag \\
\left\vert \Psi _{2}^{\left( +\right) }\right\rangle &=&N_{1}(\frac{%
E_{3}^{2}-1}{g}\left\vert \chi _{1}^{\left( +\right) }\right\rangle -\sqrt{2}%
E_{3}\left\vert \chi _{2}^{\left( +\right) }\right\rangle +\frac{E_{3}}{%
g^{\prime }}\frac{E_{5}^{2}-g^{\prime 2}}{g}\left\vert \chi _{3}^{\left(
+\right) }\right\rangle +\sqrt{2}\left\vert \chi _{4}^{\left( +\right)
}\right\rangle )\text{,} \\
\left\vert \Psi _{3}^{\left( +\right) }\right\rangle &=&N_{2}(\frac{%
E_{5}^{2}-1}{g}\left\vert \chi _{1}^{\left( +\right) }\right\rangle +\sqrt{2}%
E_{5}\left\vert \chi _{2}^{\left( +\right) }\right\rangle -\frac{E_{5}}{%
g^{\prime }}\frac{E_{3}^{2}-g^{\prime 2}}{g}\left\vert \chi _{3}^{\left(
+\right) }\right\rangle +\sqrt{2}\left\vert \chi _{4}^{\left( +\right)
}\right\rangle )\text{,}  \notag \\
\left\vert \Psi _{4}^{\left( +\right) }\right\rangle &=&N_{2}(\frac{%
E_{5}^{2}-1}{g}\left\vert \chi _{1}^{\left( +\right) }\right\rangle -\sqrt{2}%
E_{5}\left\vert \chi _{2}^{\left( +\right) }\right\rangle +\frac{E_{5}}{%
g^{\prime }}\frac{E_{3}^{2}-g^{\prime 2}}{g}\left\vert \chi _{3}^{\left(
+\right) }\right\rangle +\sqrt{2}\left\vert \chi _{4}^{\left( +\right)
}\right\rangle )\text{,}  \notag
\end{eqnarray}%
where $N_{1}$ and $N_{2}$ are normalized factors.

Now, the dynamical evolution of the initial state%
\begin{eqnarray}
\left\vert d\left( 0\right) \right\rangle &=&N(g\left\vert 1\right\rangle
_{I}\left\vert 0\right\rangle _{II}\left\vert 0\right\rangle
_{c}+g\left\vert 0\right\rangle _{I}\left\vert 1\right\rangle
_{II}\left\vert 0\right\rangle _{c}-\left\vert 0\right\rangle _{I}\left\vert
0\right\rangle _{II}\left\vert 1\right\rangle _{c})\left\vert g\right\rangle
_{A}  \notag \\
&=&N\left( \sqrt{2}g\left\vert \chi _{4}^{\left( +\right) }\right\rangle
-\left\vert \chi _{1}^{\left( +\right) }\right\rangle \right) \text{,}
\label{d0}
\end{eqnarray}%
where $N=1/\sqrt{2g^{2}+1}$ normalized factor, can be calculated. From (\ref%
{d0}) we find that $\left\vert d\left( 0\right) \right\rangle $ is in the
symmetrical subspace. At time $t$, it evolves into%
\begin{eqnarray}
\left\vert d\left( t\right) \right\rangle &=&F_{1}\left( t\right) \left\vert
\chi _{1}^{\left( +\right) }\right\rangle +F_{2}\left( t\right) \left\vert
\chi _{2}^{\left( +\right) }\right\rangle +F_{3}\left( t\right) \left\vert
\chi _{4}^{\left( +\right) }\right\rangle +F_{4}\left( t\right) \left\vert
\chi _{3}^{\left( +\right) }\right\rangle  \notag \\
&=&F_{1}\left( t\right) \left\vert 0\right\rangle _{I}\left\vert
0\right\rangle _{II}\left\vert 1\right\rangle _{c}\left\vert g\right\rangle
_{A}+F_{2}\left( t\right) \left\vert D\right\rangle \left\vert
0\right\rangle _{c}\left\vert g\right\rangle _{A} \\
&&+F_{3}\left( t\right) \left\vert C\right\rangle \left\vert 0\right\rangle
_{c}\left\vert g\right\rangle _{A}+F_{4}\left( t\right) \left\vert
0\right\rangle _{I}\left\vert 0\right\rangle _{II}\left\vert 0\right\rangle
_{c}\left\vert e\right\rangle _{A}\text{,}  \notag
\end{eqnarray}%
where%
\begin{eqnarray}
F_{1}\left( t\right) &=&2N\left( t\right) [N_{1}^{2}\frac{E_{3}^{2}-1}{g}%
X\cos \left( E_{3}t\right) +Z\frac{E_{5}^{2}-1}{g}\cos \left( E_{5}t\right) ]%
\text{,}  \notag \\
F_{2}\left( t\right) &=&-i2\sqrt{2}N\left( t\right) [N_{1}^{2}E_{3}X\sin
\left( E_{3}t\right) +ZE_{5}Y\sin \left( E_{5}t\right) ]\text{,}  \notag \\
F_{3}\left( t\right) &=&2\sqrt{2}N\left( t\right) [N_{1}^{2}X\cos \left(
E_{3}t\right) +ZY\cos \left( E_{5}t\right) ]\text{,} \\
F_{4}\left( t\right) &=&i2N\left( t\right) [\frac{N_{1}^{2}E_{3}}{g^{\prime }%
}\frac{E_{5}^{2}-g^{\prime 2}}{g}X\sin \left( E_{3}t\right) +ZE_{5}\frac{%
E_{3}^{2}-g^{\prime 2}}{gg^{\prime }}Y\sin \left( E_{5}t\right) ]\text{,}
\notag
\end{eqnarray}%
and%
\begin{eqnarray}
X &=&2g-\frac{E_{3}^{2}-1}{g}\text{,}  \notag \\
Y &=&2g-\frac{E_{5}^{2}-1}{g}\text{,}  \notag \\
Z &=&\frac{1}{4}-N_{1}^{2}\text{,} \\
\left\vert D\right\rangle &=&\frac{1}{\sqrt{2}}\left( \left\vert
a\right\rangle _{I}\left\vert 0\right\rangle _{II}+\left\vert 0\right\rangle
_{I}\left\vert a\right\rangle _{II}\right) \text{,}  \notag \\
\left\vert C\right\rangle &=&\frac{1}{\sqrt{2}}\left( \left\vert
1\right\rangle _{I}\left\vert 0\right\rangle _{II}+\left\vert 0\right\rangle
_{I}\left\vert 1\right\rangle _{II}\right) \text{,}  \notag
\end{eqnarray}%
where $N\left( t\right) $ is the normalized factor.

Some observations follow: (1) States $\left\vert g\right\rangle _{A}$\ and $%
\left\vert e\right\rangle _{A}$ of the auxiliary SQUID are entangled with
states of two SQUIDs and the cavity field; (2) When the measurement to the
auxiliary SQUID is carried out, the auxiliary SQUID will collapse into a
quantum state. If the auxiliary SQUID collapses to the ground state $%
\left\vert g\right\rangle _{A}$, one can not determine which state two
SQUIDs will be in. This is because there exist three states ($\left\vert
0\right\rangle _{I}\left\vert 0\right\rangle _{II}$, $\left\vert
C\right\rangle $, and $\left\vert D\right\rangle $) of two SQUIDs entangled
with state $\left\vert g\right\rangle _{A}$. In the following section, we
will discuss how to create the entangled state $\left\vert C\right\rangle $
by choosing optimal parameters.

\section{IV. Creating maximally entangled state by optimizing parameters}

There exist two entangled states $\left\vert C\right\rangle $ and $%
\left\vert D\right\rangle $ in state $\left\vert d\left( t\right)
\right\rangle $. In this paper, $\left\vert C\right\rangle $ is the target
state in accordance with Ref. \cite{Yang}. The entangled state $\left\vert
D\right\rangle $ contains excited states $\left\vert a\right\rangle _{l}$ of
two SQUIDs, which couple to the environment with a higher probability of
transitions to the first excited state $\left\vert 1\right\rangle _{l}$ or
the ground state $\left\vert 0\right\rangle _{l}$, so state $\left\vert
D\right\rangle $ is unstable. Therefore, state $\left\vert C\right\rangle $
is a good candidate for quantum information process.

To create entangled state $\left\vert C\right\rangle $, we first choose a
special instant $t_{0}$ to satisfy%
\begin{equation}
P_{1}\left( t_{0}\right) =P_{2}\left( t_{0}\right) =0\text{,}
\label{entanglement}
\end{equation}%
where $P_{1}\left( t_{0}\right) =\left\vert F_{1}\left( t_{0}\right)
\right\vert ^{2}$ and $P_{2}\left( t_{0}\right) =\left\vert F_{2}\left(
t_{0}\right) \right\vert ^{2}$. The above equations can be regarded
\textquotedblleft entanglement condition\textquotedblright , which introduce
the relation between $E_{3}$ and $E_{5}$. The parameters can be determined
when $P_{3}\left( t_{0}\right) =\left\vert F_{3}\left( t_{0}\right)
\right\vert ^{2}$, which is the probability of state $\left\vert
C\right\rangle \left\vert 0\right\rangle _{c}\left\vert g\right\rangle _{A}$%
, reaches its maximum. Then the entangled state $\left\vert C\right\rangle $
can be generated with higher success probability. Based on the entanglement
condition (\ref{entanglement}), numerical simulations are employed to find
the maximal value of $P_{3}\left( E_{3}\right) $. Numerical results are
listed in Figs. 3, 4.

\begin{figure}[tbp]
\includegraphics[width=15 cm]{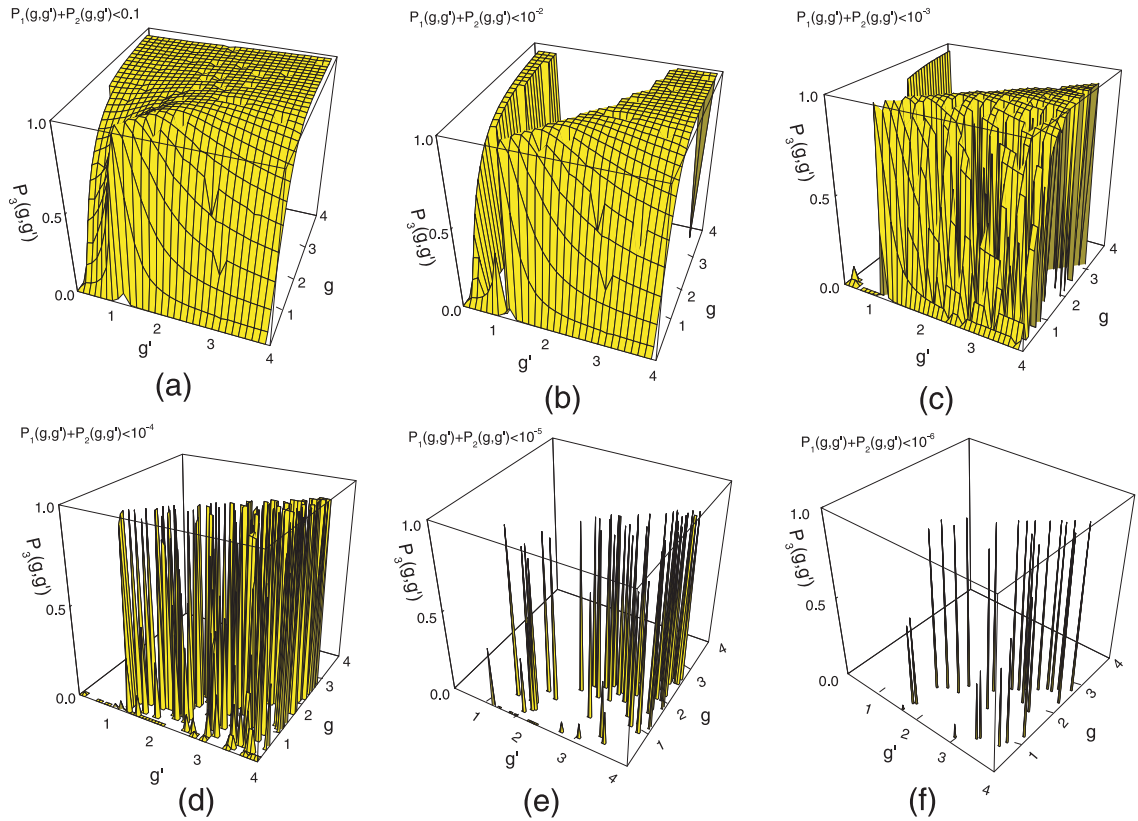}
\caption{\textit{The relation between maximal values of $P_{3}$ and $g$ as
well as $g^{\prime }$ when $P_{1}+P_{2}<10^{-i}$($i=1,2,...6$). The
influence of the auxiliary SQUID is taken into account as a quantum
subsystem. As a result the parameter $g$ of two SQUIDs and the parameter $%
g^{\prime }$ of the auxiliary SQUID must match so that the target entangled
state can be induced. Numerical results show that the closer to the
entanglement condition are the results the less is the number of $g$ and $%
g^{\prime }$.}}
\end{figure}

As shown in Fig. 3, when $P_{1}\left( t_{0}\right) +P_{2}\left( t_{0}\right)
$ is less than $10^{-j}$ ($j=1\sim 6$), the maximal values of $P_{3}\left(
t_{0}\right) $ and magnitudes of $g$ and $g^{\prime }$ corresponding to
these $P_{3}\left( t_{0}\right) $ are given respectively. Here, the
influence of the auxiliary SQUID is taken into account as a quantum
subsystem. As a result the parameter $g$ of the two SQUIDs and the parameter
$g^{\prime }$ of the auxiliary SQUID must match so that the entangled state
we need can be induced. The numerical results show that as $P_{1}\left(
t_{0}\right) +P_{2}\left( t_{0}\right) $ approaches to vanishing, the
optimal area in $gg^{\prime }$ plane, within which $P_{3}\left( t_{0}\right)
$ closes to unit.

\begin{figure}[tbp]
\includegraphics[width=15 cm]{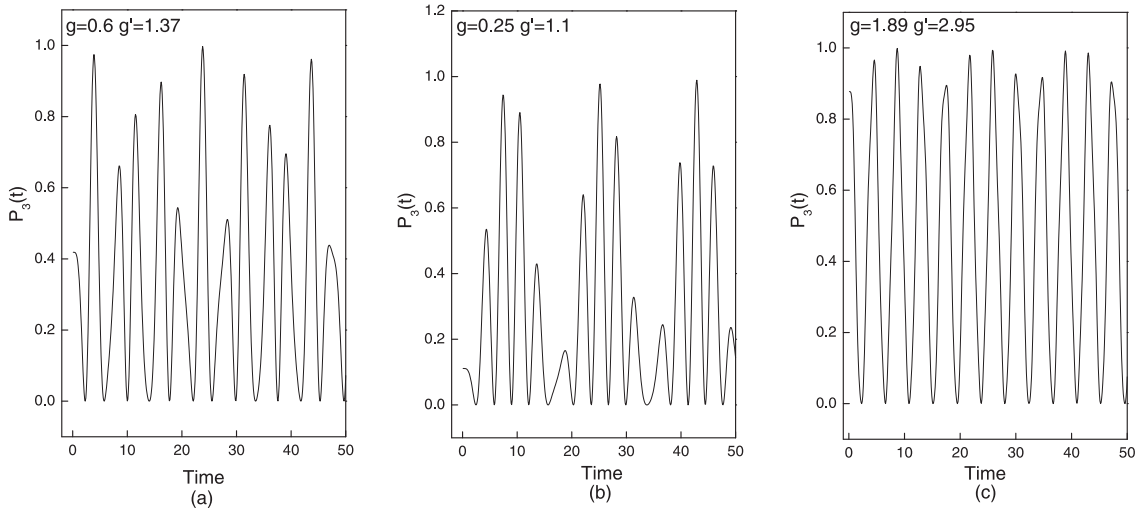}
\caption{\textit{The relation between time and $P_{3}\left( t\right) $. In
the case of $P_{1}\left( t_{0}\right) +P_{2}\left( t_{0}\right) $ less than $%
10^{-6}$, three groups $g$ and $g^{\prime }$ are chosen. Numerical results
show that $t_{0}$ which make $P_{3}\left( t_{0}\right) $ maximal is not $%
\protect\pi /g^{\prime }$ due to the back reaction of the auxiliary SQUID to
two SQUIDs.}}
\end{figure}

We also investigate the dynamical process of the entanglement creation by
numerical simulation. For the case with fixed $g$ and $g^{\prime }$, under
the condition\ $P_{1}\left( t_{0}\right) +P_{2}\left( t_{0}\right) \leq
10^{-6}$, the time evolution of $P_{3}\left( t\right) $ is calculated.

Three groups of $g\ $and $g^{\prime }$ with $(g,\ g^{\prime })=(0.25,$ $%
1.89),$\ $(2.95,$ $1.10),$\ $(0.60,$ $1.37)$\ are chosen to demonstrate the
dynamical process of the creation of entanglement. Corresponding initial
states are $\left\vert d\left( 0\right) \right\rangle =$ $N\left( \sqrt{2}%
g\left\vert \chi _{4}^{\left( +\right) }\right\rangle -\left\vert \chi
_{1}^{\left( +\right) }\right\rangle \right) $\ with $N=1/\sqrt{2g^{2}+1}$.
In each case, the component of maximal entangled state in the initial state
is $2g^{2}/N^{2}$. Numerical results are plotted in Fig. 4. Taking Fig. 4(b)
as an example, the initial state is the dark state of the two SQUIDs, which
contains a small component ($\sim 0.11$) of the maximal entangled state.
When the coupling to the auxiliary SQUID is switched on, i.e., $g^{\prime
}=0\longrightarrow 1.10$, to probe the state of the cavity field. After $%
t_{0}$, the quantum state which contains a maximal component of state $%
\left\vert C\right\rangle $ is generated. Then the post selection
measurement to the auxiliary SQUID is carried out. As a result of maximizing
the probability $P_{3}\left( t_{0}\right) $, there exits very large
probability to induce the state $\left\vert C\right\rangle $. This can also
be found from Fig. 4(a,c). It indicates that the smaller component of state $%
\left\vert C\right\rangle $ is contained in the initial state, the longer
time period $t_{0}$ is taken. Numerical results also show that period $t_{0}$%
\ at which $P_{3}\left( t_{0}\right) $ gets maximum is no longer $\pi
/g^{\prime }$ due to the back reaction of the auxiliary SQUID to the two
SQUIDs.

\section{V. Summary}

In this paper, the back action of the auxiliary SQUID to the two SQUIDs was
taken into account in the theoretical scheme for creating the entangled
state. The dynamical evolution of the two SQUIDs dark state (\ref{dark})
driven by the Hamiltonian of two SQUIDs together with the auxiliary SQUID is
calculated precisely. By using the entanglement condition, the relation
between parameters ($g$ and $g^{\prime }$) and the evolution time $t_{0}$ is
determined. At instant $t_{0}$, the system reaches state $\left\vert d\left(
t\right) \right\rangle =F_{3}\left( t\right) \left\vert C\right\rangle
\left\vert 0\right\rangle _{c}\left\vert g\right\rangle _{A}+F_{4}\left(
t\right) \left\vert 0\right\rangle _{I}\left\vert 0\right\rangle
_{II}\left\vert 0\right\rangle _{c}\left\vert e\right\rangle _{A}$. By
optimizing the parameters ($g$ and $g^{\prime }$), the probability $%
P_{3}\left( t_{0}\right) =\left\vert F_{3}\left( t_{0}\right) \right\vert
^{2}$ of the entangled state $\left\vert C\right\rangle $ is maximized at $%
t_{0}$. After the post selection measurement for the auxiliary SQUID,
entangled state $\left\vert C\right\rangle $ is feasibly induced. Numerical
results show that the period $t_{0}$ given by the entanglement condition is
not $\pi /g^{\prime }$. This is because the auxiliary SQUID is taken into
account as a quantum subsystem.

We gratefully acknowledge the valuable discussion with Professor Chang-Pu
Sun. This work is supported by the CNSF (Grant No. 10474104), the National
Fundamental Research Program of China (No. 2001CB309310).


\end{document}